\documentclass[conference]{IEEEtran}
\IEEEoverridecommandlockouts

\usepackage{graphicx} 
\usepackage{pdfpages}
\usepackage{multirow, makecell} 
\usepackage{booktabs}
\usepackage{hyperref}
\usepackage{multirow}
\usepackage{multicol}
\usepackage{float}
\usepackage{amssymb}
\usepackage{csquotes}
\usepackage[figuresright]{rotating}
\usepackage{tablefootnote}
\usepackage{xcolor,colortbl}
\usepackage{threeparttable}
\usepackage{geometry}
\begin{document}

\title{Cultural Impact on Requirements Engineering Activities: Bangladeshi
Practitioners’ View}


\author{\IEEEauthorblockN{Chowdhury Shahriar Muzammel}
\IEEEauthorblockA{\textit{RMIT University, Australia}\\
s3987367@student.rmit.edu.au}
\and
\IEEEauthorblockN{Maria Spichkova}
\IEEEauthorblockA{\textit{RMIT University, Australia}\\
maria.spichkova@rmit.edu.au}
\and
\IEEEauthorblockN{James Harland}
\IEEEauthorblockA{\textit{RMIT University, Australia}\\
james.harland@rmit.edu.au}
}
\maketitle



\section{Context}

Requirements Engineering (RE) is one of the most interaction-intensive phases of software development. 
This means that RE activities might be especially impacted by stakeholders' national culture. Software development projects increasingly have a very diverse range of stakeholders. To future-proof RE activities, we need to help RE practitioners avoid misunderstandings and conflicts that might arise from not understanding
potential \emph{Cultural Influences} (CIs). Moreover, an awareness of CIs supports diversity and inclusion in the IT profession. Bangladesh has a growing IT sector with some unique socio-cultural characteristics, and has been largely overlooked in this research field. In this study, we aim to investigate how the RE process is adopted in the context of Bangladeshi culture and what cultural influences impact overall RE activities. 
\\
~\\
\emph{Preprint. Accepted to the 33rd IEEE International Requirements Engineering Conference (RE'25), September 1–5, 2025, Valencia, Spain. IEEE Xplore. Final version to be published by IEEE (In Press).}

\section{RE problem and Motivation}
We follow the definition of \emph{culture} introduced by Hofstede et al.~\cite{hofstede2010cultures}: Culture means how we behave, act/react, think, understand, make decisions, and communicate in a society with each other. Every nationality has distinct beliefs, customs, and communication processes, which can be identified as national culture. National culture might even dominate our human communication in RE activities, which impacts the overall process. 
Therefore, it's important to study a country's culture before starting RE activities.

As the Bangladeshi software industry is growing rapidly and exporting worldwide, it is important to understand the RE process and related challenges in Bangladeshi culture. To date, to the best of our knowledge, no case study has been available regarding cultural impact on RE activities in the Bangladeshi context. Thus, we define the research question for this study as follows: 
\emph{\textbf{RQ:} What cultural influences impact RE activities in Bangladesh?}

The results of the study can be directly used by practitioners. 
However, by answering the above research question, we also aim to investigate whether a broader solution might be possible: we aim to extend an existing framework by Alsanoosy et al.~\cite{alsanoosy2020identification} to cover the countries where no primary case studies have been conducted previously. 

\section{Methodology}

We conducted a case study with Bangladeshi practitioners following the qualitative research method.  We employed  semi-structured interviews, which involve asking predefined open-ended questions to gather extensive insights from experts. We followed the steps below:

\textbf{Step 1: Data Collection.} We have created semi-structured interview questions to fit our research scope for Bangladesh. We recruited participants by examining their backgrounds and reviewing websites from relevant software companies. 
We interviewed twenty Bangladeshi software practitioners involved in RE-related activities across various software development companies to explore the national CIs affecting RE-related activities. Table~\ref{tab:ParticipantsInfo} summarises the demographical data of the participants, where \emph{Y} denotes years of software engineering experience, \emph{Size} denotes the company size, \emph{A/W} denotes the methodology used (Agile/Waterfall), and \emph{P/G} denotes the company type (Private/Governmental).

 \renewcommand{\arraystretch}{0.8}
\begin{table}[ht]
\centering
\begin{footnotesize}
\caption{Demographic data of study participants}
\label{tab:ParticipantsInfo}
\begin{tabular}{l|l|l|l|l|l}
\hline
\textbf{ID}  & \textbf{Y}  & \textbf{Role} & \textbf{Size} & \textbf{A/W} & \textbf{G/P} \\      \hline \hline
    
       P1  & 6 & Software Eng./ Req. Eng. & L & A & P \\ \hline
       P2  & 4 & Software Eng./ Req. Eng. & M & W & P \\ \hline
       P3  & 3 & Software Eng./ Req. Eng. & M & W & G/ P \\ \hline
       P4  & 7 & System Analyst/ Req. Eng. & M & A & G/ P \\ \hline
       P5  & 5 & Software Dev./ Req. Eng. & M & A & P \\ \hline
       P6  & 3 & Software Eng./ Req. Eng. & L & A & P \\ \hline
       P7  & 4 & Software Dev./ Req. Eng. & M & A, W & P \\ \hline
       P8  & 4 & Software Eng./ Req. Eng. & M & W & P \\ \hline
       P9  & 8 & Data Analyst/ Req. Eng. & L & A & P \\ \hline
       P10  & 6 & Software Eng./ Req. Eng. & L & A & P \\ \hline
       P11  & 7 & Team Lead/ Req. Eng. & M & A & P \\ \hline
       P12  & 3 & Software Eng./ Req. Eng. & L & A & P \\ \hline
       P13  & 5 & Sales Eng./ Req. Eng. & M & A & G/ P \\ \hline
       P14  & 6 & Req. Eng. & M & A & G/ P \\ \hline
       P15  & 8 & Product Manager/ Req. Eng. & L & A & P \\ \hline
       P16  & 5 & Project Manager/ Req. Eng. & L & A & P \\ \hline
       P17  & 5 & Software Dev./ Req. Eng. & L & A & P \\ \hline
       P18  & 3 & System Dev./ Req. Eng. & M & A & P \\ \hline
       P19  & 3 & System Dev./ Req. Eng. & M & W & P \\ \hline
       P20  & 6 & Team Lead/ Req. Eng. & L & A & P \\ \hline
\end{tabular}
\end{footnotesize}
\end{table}

\renewcommand{\arraystretch}{0.85}
\begin{table*}[ht]
\centering
\begin{footnotesize}
\caption{Summary of Interview Results}
\label{tab:ResultSummary}
\begin{tabular}{l|c|c|c|l|c}
\hline
\textbf{Cultural influences}  & \textbf{Yes}  & \textbf{Positive }& \textbf{Negative} & \textbf{RE activities} & \textbf{Identified by} \\ & & & & \textbf{affected}  & \textbf{framework}\\      \hline \hline
                     
CI.1) Centralised decision-making & 17 & 2 & 15 & S V M & \checkmark \\ \hline 
CI.2) Communication context & 17 & - & 17 & E V M & $\times$ \\ \hline
CI.3) Managers’ influence & 11 & - & 11 & All & \checkmark \\ \hline
CI.4) Clients’ resistance & 11 & - & 11 & S V M & \checkmark \\ \hline
CI.5) Coordination and collaboration/Teamwork & 9 & - & 9 & All & $\times$\\ \hline
CI.6) Aiming for quick results & 7 & - & 7 & All & \checkmark \\ \hline
CI.7) Deference & 6 & - & 6 & E A S & \checkmark\\ \hline
CI.8) Establish trust & 6 & - & 6 & E V M & \checkmark\\ \hline
CI.9) Punctuality & 6 & - & 6 & E S M & $\times$ \\ \hline
CI.10) Language/Accent & 6 & - & 6 & E V M & $\times$ \\ \hline
CI.11) Solving conflicts by favoritism & 5 & - & 5 & S V M & \checkmark\\ \hline
CI.12) Employees' attitude & 4 & - & 4 & E S M & $\times$ \\ \hline
CI.13) Subordinate avoid taking risks & 4 & - & 4 & S V & \checkmark \\ \hline
CI.14) Gender preference & 4 & - & 4 & E A S M & $\times$ \\ \hline
CI.15) Loose employment of RE practice & 3 & - & 3 & All & $\times$ \\ \hline
CI.16) Building relationships & 10 & 10 & - & E A V M & \checkmark \\ \hline 
CI.17) Relying on previous   projects & 3 & 3 & - & All & \checkmark\\ \hline
CI.18) Avoiding conflicts & 1 & - & 1 & All & \checkmark \\ \hline

\end{tabular}
\end{footnotesize}
\end{table*}

\textbf{Step 2: Data Analysis.} We used thematic analysis (TA) to derive insights from our collected interviews. All interviews were conducted in Bengali for the participants' convenience and preference. The sessions were recorded to facilitate later translation into English. Each interview was manually translated and transcribed by the first author.   
To follow TA, we went through the following steps proposed by Braun and Clarke~\cite{braun2006using}.
The first author read the data to become familiar with the data, and conducted open coding to thoroughly analyse the data. Open coding can help the researcher to identify the main concerns raised by participants~\cite{hoda2017becoming}. Once all \emph{codes} and \textit{themes} were identified, the second and third authors reviewed them to validate the results. Next, several meetings were organised to finalise the \emph{codes}  and \textit{themes}.

    \textbf{Step 3: Framework vs. Case Study.} 
    We compared the results of our study with the results generated for Bangladesh by the framework proposed by Alsanoosy et al.~\cite{alsanoosy2020identification}. 
    The aim of this comparison was to identify 
    potential future improvements of the framework.

\section{Results and Contributions}

Table~\ref{tab:ResultSummary} provides an overview of the results of our study.
Overall, we identified 18 CIs. We indicated how frequently practitioners reported each CI, whether the practitioners perceived them as having a positive or negative impact, as well as what RE activity, i.e., Elicitation (E), Analysis (A), Specification (S), Validation (V), and
Management (M), might be affected by it from the practitioner's view.  While some CI have been mentioned by the majority of the participants with unanimous labelling of this CI as positive/negative (e.g., \emph{Managers' influence} was mentioned by 11 participants and all of them perceived it as negatively impacting RE activities), other CIs have been mentioned only a small number of participants, which demonstrates the need to clarify this aspect further. Only for one of the identified CIs, \emph{Centralised decision-making}, the perceptions regarding its impact weren't unanimous: 15 participants perceived it as having a negative impact (e.g., P19 explained it by \emph{"If its centralised, then we don’t get the decision in time, it gets delayed,"} while 2 participants viewed it as having a positive impact, assuming that it is based on relevant technical knowledge.

We compared the identified CIs with the anticipated outcome for Bangladesh generated by the framework~\cite{alsanoosy2020identification}.
All of the CIs generated by the framework were also identified by our study. However, the framework generated only 11 out of 18 CIs (approx. \textit{61\%}) explored by the current case study.
Thus, the framework generally provides potentially helpful results, but it would be useful to refine it further to provide better coverage. 
At present, the primary case study conducted with Bangladeshi practitioners provides a more detailed overview of the cultural influences.

\section{Related Work}
There are several studies on the impact of national culture on the RE activities.
Kanij et al.~\cite{kanij2023developing} investigated specifics of collecting requirements from low socio-economic end users, such as Bangladeshi fisherfolk. 
Rahim et al.~\cite{rahim2017software} conducted a survey on Bangladeshi software engineering practices and challenges. The survey indicated the major challenges are related to proper RE and its prioritising, decision-making process, as well as a communication gap between clients and stakeholders.  
This highlights that taking care of national culture is critical while developing software in the Bangladeshi context. 

As a groundwork to refine the framework, we conducted a systematic literature review (SLR)~\cite{muzammel2024cultural}, which summarised the latest list of CIs impacting RE-related activities.  
As our future work, we plan to refine the framework based on the SLR as well as the results of the presented study.

\bibliographystyle{IEEEtran}
\bibliography{sources}

\end{document}